\documentclass[10pt,letterpaper]{article}
\usepackage[top=0.85in,left=2.75in,footskip=0.75in]{geometry}

\usepackage{amsmath,amssymb}

\usepackage{changepage}

\usepackage{textcomp,marvosym}

\usepackage{cite}

\usepackage{nameref,hyperref}

\usepackage[nopatch=eqnum]{microtype}
\DisableLigatures[f]{encoding = *, family = * }

\usepackage[table]{xcolor}

\usepackage{array}

\newcolumntype{+}{!{\vrule width 2pt}}

\newlength\savedwidth

\usepackage{setspace} 
\raggedright
\usepackage[aboveskip=1pt,labelfont=bf,labelsep=period,justification=raggedright,singlelinecheck=off]{caption}

\makeatletter
\renewcommand{\@biblabel}[1]{\quad#1.}
\makeatother

\usepackage{lastpage,fancyhdr,graphicx}
\usepackage{epstopdf}
\fancyheadoffset[L]{2.25in}
\fancyfootoffset[L]{2.25in}
\usepackage{amsfonts}
\usepackage{booktabs}
\usepackage{multirow}
\usepackage{subcaption}
\usepackage{algorithm}
\usepackage{algpseudocode}
\usepackage{float}
\usepackage{enumitem}
\usepackage[nopatch=eqnum]{microtype}
\begin{document}
\vspace*{0.2in}

% Title must be 250 characters or less.
\begin{flushleft}
{\Large
\textbf\newline{Binary Gaussian Copula Synthesis: an LLM-powered data augmentation framework for early dialysis prediction in 
chronic kidney disease}
}
\newline
% Short title must be 70 characters or less
{\textit{Short title: }LLM-augmented framework for early dialysis prediction in CKD}
\newline
% Insert author names, affiliations and corresponding author email (do not include titles, positions, or degrees).
\\
Hamed Khosravi\textsuperscript{1,6},
Milad Khanchi\textsuperscript{2\Yinyang},
Mobina Noori\textsuperscript{3\Yinyang},
Srinjoy Das\textsuperscript{4},
Mohammad Abdullah Al-Mamun\textsuperscript{5},
Imtiaz Ahmed\textsuperscript{1*}

\bigskip
\textbf{1} Department of Industrial \& Management Systems Engineering, West Virginia University, Morgantown, WV, USA
\\
\textbf{2} Department of Electrical and Computer Engineering, Concordia University, Montreal, QC, Canada
\\
\textbf{3} Department of Computer Science, University of California, Davis, CA, USA
\\
\textbf{4} School of Mathematical \& Data Sciences, West Virginia University, Morgantown, WV, USA
\\
\textbf{5} School of Systems Science and Industrial Engineering, The State University of New York at Binghamton, Binghamton, NY, USA
\\
\textbf{6} H.~Milton Stewart School of Industrial and Systems Engineering, Georgia Institute of Technology, Atlanta, GA, USA
\\
\bigskip

% Primary Equal Contribution Note
\Yinyang These authors contributed equally to this work.

% Use the asterisk to denote corresponding authorship and provide email address in note below.
* imtiaz.ahmed@mail.wvu.edu

\end{flushleft}

% Please keep the abstract below 300 words
\section*{Abstract}
Only a small fraction of patients with chronic kidney disease (CKD) progress to dialysis, creating severe class imbalance that limits the performance of machine learning models for early dialysis prediction. This challenge is compounded by the binary structure of electronic health record (EHR) data, for which most existing augmentation methods were not designed. We propose Binary Gaussian Copula Synthesis (BGCS), a two-stage data augmentation method tailored to binary clinical data. BGCS first generates synthetic minority-class samples using a Gaussian copula framework that explicitly models pairwise dependencies among binary features, then applies a fine-tuned GPT-2 classifier to filter out clinically implausible samples before training. We evaluated BGCS on a real-world EHR dataset of 15,169 patients with CKD from West Virginia collected between 2008 and 2022, benchmarking it against SMOTE, CTGAN, and standard Gaussian Copula across four machine learning classifiers over 25 independent runs. BGCS consistently outperformed all comparison methods, achieving the highest minority-class recall for 90-day dialysis prediction, with median values ranging from 0.78 to 0.87 across classifiers, and the strongest distributional fidelity to real data, with a mean p-value of 0.68 across features. The best-performing BGCS-augmented model was integrated into an interpretable decision tree–based clinical decision support system for dialysis risk stratification, with electrolyte imbalances, cardiovascular comorbidities, and renal monitoring indicators emerging as the most influential predictive features. These findings suggest that augmentation methods designed for the structural properties of binary EHR data can meaningfully improve early dialysis risk prediction and support the development of interpretable clinical decision-support tools for CKD care.

% Please keep the Author Summary between 150 and 200 words
% Use first person. Do not use subheadings for PLOS Digital Health.

\section*{Author summary}

Many patients with chronic kidney disease are not recognized as being at high risk for dialysis until the disease reaches an advanced stage. Earlier prediction could give clinicians more time to intervene and help patients prepare for treatment. This is challenging for two reasons. First, the number of patients who eventually require dialysis is much smaller than the number who do not, which makes model training and accurate prediction difficult. Second, many clinical records are stored as yes-or-no variables, and existing methods do not handle this type of data well. In this study, we developed a new method called Binary Gaussian Copula Synthesis (BGCS) to create realistic synthetic patient records from this kind of clinical data. We used these records to help prediction models better identify patients at risk of future dialysis. Using electronic health records from more than 15,000 patients with chronic kidney disease in West Virginia, we found that BGCS performed better than existing approaches. We then used the best-performing model to build a clinical decision support tool that helps identify patients at elevated risk of dialysis and shows the clinical features associated with that risk. Overall, our findings suggest that this approach can improve early prediction and support earlier intervention.

% ==========================================
% INTRODUCTION
% ==========================================
\section*{Introduction}

Chronic kidney disease (CKD) is a major global health challenge. Over 37 million adults in the United States and more than 800 million individuals worldwide live with CKD, which has an estimated prevalence exceeding 10\%~\cite{kovesdy2022epidemiology}. In the United States alone, treatment costs for Medicare beneficiaries with CKD and ESRD reached \$87.2 billion and \$37.3 billion, respectively. Over the past two decades, CKD has become one of the few non-communicable diseases with rising mortality rates~\cite{gembillo2023lung}. Because CKD is largely asymptomatic in its early stages, diagnosis is frequently delayed, limiting opportunities for preventative intervention and increasing the risk of irreversible kidney damage~\cite{silveira2022exploring}. Consequently, the number of patients requiring dialysis has tripled over the past decade~\cite{ahn2022factors}, and more than 808,000 individuals in the United States currently depend on dialysis or kidney transplantation. These clinical realities also create an important methodological challenge. Although early identification of patients at risk of dialysis is clinically important, only a small subset of patients with CKD ultimately progress to dialysis. As a result, this classification task involves a substantial imbalance between patients who do and do not progress to dialysis.

Imbalanced class distributions represent a significant challenge in data science and machine learning (ML), particularly in classification tasks where some classes have far fewer examples than others~\cite{lee2022downsampling, alam2022efficient}. Because standard classifiers tend to favor majority classes~\cite{galar2011review} and often treat minority-class instances as noise~\cite{islam2022knnor}, most ML methods perform worse under such conditions~\cite{gao2020handling}, and minority-class samples are more likely to be misclassified~\cite{japkowicz2002class, vandewiele2021overly}. Class imbalance is prevalent across many real-world applications, including medical diagnosis, spam detection~\cite{liu2017addressing}, credit card fraud detection~\cite{dhankhad2018supervised}, and industrial failure prediction~\cite{lee2021early}. Medical datasets are particularly vulnerable to this problem, often exhibiting marked differences in the frequency of common and rare conditions~\cite{kumar2022addressing, saarela2019predicting}, notably within electronic health records (EHRs)~\cite{khushi2021comparative, salau2023influence}. In these settings, standard classifiers frequently produce biased predictions that favor the majority class~\cite{norori2021addressing, yang2023algorithmic}, compromising model reliability and limiting the detection of clinically important minority cases. This challenge is highly relevant in CKD, where only a small fraction of patients progress to dialysis, creating pronounced class imbalance and making early and accurate prediction more difficult~\cite{vadakedath2017dialysis}.

Given this challenge, a range of resampling strategies has been proposed to address class imbalance~\cite{liu2022solving, cohen2006learning}. Random oversampling and undersampling are the simplest approaches, but they may duplicate non-informative minority observations or remove potentially informative majority-class samples~\cite{wegier2020application}. More sophisticated methods, including the Synthetic Minority Over-sampling Technique (SMOTE)~\cite{chawla2002smote}, Adaptive Synthetic Sampling (ADASYN)~\cite{he2008adasyn}, and Majority Weighted Minority Oversampling Technique (MWMOTE)~\cite{barua2012mwmote}, create synthetic minority-class samples from existing data, although they differ in how those samples are selected and generated. These methods, particularly SMOTE and its variants, have been applied widely in healthcare prediction studies. Prior work has compared SMOTE with random undersampling for post-bariatric surgery complication prediction~\cite{razzaghi2019predictive}, integrated Modified SMOTE with Edited Nearest Neighbors in hybrid preprocessing frameworks~\cite{xu2020hybrid}, and applied SMOTE-based resampling to CKD progression modeling~\cite{shi2022resampling}. Similar resampling strategies have also been explored in trauma-related datasets~\cite{hassanzadeh2023hospital}, mechanically ventilated patient cohorts~\cite{hirano2021machine}, and chronic non-communicable disease datasets evaluating SMOTE and ADASYN~\cite{rodriguez2022synthetic}. However, the suitability of many commonly used resampling methods for predominantly binary EHR data remains uncertain, particularly when synthetic samples are generated through interpolation rather than from an explicit model of dependence.

To move beyond interpolation-based resampling, recent studies have explored more flexible synthetic data generation approaches. Deep generative models such as generative adversarial networks (GANs)~\cite{mirza2014conditional, berger2023generative} and copula-based statistical methods~\cite{xue2000multivariate} have broadened the set of available methods for handling imbalanced clinical data. Prior work has applied these approaches to hybrid preprocessing for stroke prediction~\cite{liu2019hybrid}, VAE-GAN models for clinical data synthesis~\cite{nicholas2023generating}, clustering-based methods for emergency department outcomes~\cite{chen2023dealing}, and joint frameworks for missing-value imputation and imbalanced learning in EHRs~\cite{weng2024joint}. However, many of these methods do not explicitly model how binary clinical variables co-occur, which may limit the realism of the synthetic data they generate. This limitation is important in EHR-based studies, where diagnosis and procedure codes are typically recorded as present-or-absent indicators~\cite{zhou2023approaches}. 

Appropriate evaluation is equally important in imbalanced settings. In these contexts, standard performance measures can be misleading because high overall accuracy may be achieved simply by predicting the majority class, a phenomenon often referred to as the accuracy paradox~\cite{valverde2014100, vadakedath2017dialysis}. For clinical applications such as dialysis prediction, recall is particularly important because it reflects the proportion of true positive cases correctly identified, and failing to detect a patient who will require dialysis may have serious consequences~\cite{zhao2018framework, hicks2022evaluation}. Precision and the F1 score provide additional insight by helping assess the reliability and balance of model performance. This emphasis on clinically meaningful evaluation is especially important when predictive models are intended for use in decision-support tools at the point of care.

Clinical decision support systems (CDSSs) provide one pathway for translating predictive models into practice. These systems range from simple alert tools to more advanced ML and artificial intelligence (AI)-driven frameworks, all designed to support clinicians by delivering relevant information at the point of care~\cite{du2022explainable}. Their growing use in healthcare has been driven by advances in ML that have improved prediction, diagnosis, prognosis, and pattern recognition~\cite{smiti2020machine}. For CKD care, a CDSS could help identify patients at elevated risk of dialysis earlier in the disease course, enabling more timely and targeted intervention~\cite{afrash2022design}. However, the complexity of clinical data, together with practical barriers to implementation, continues to limit widespread real-world adoption.

In line with this translational goal, recent work has demonstrated growing interest in the use of ML for CKD care, with applications spanning clinical risk assessment, mortality prediction in high-risk populations, short-term prediction of kidney failure, and modeling of CKD progression~\cite{lu2023risk, ye2023prediction, klamrowski2023short, aoki2023ckd, liang2023deep}. Together, these studies highlight the promise of ML for supporting earlier and more informed clinical intervention. However, important gaps remain. Early prediction of dialysis initiation has received comparatively less attention, particularly in settings that support interpretability and clinical decision-making through a CDSS framework~\cite{hassanat2022rdpvr}. Moreover, the challenges of class imbalance in binary EHR data and the need to rigorously assess the quality of synthetic augmentation remain insufficiently addressed. These issues become paramount when predictive models are developed for use in high-stakes clinical settings.

In this study, we address these gaps through three main contributions. First, we propose Binary Gaussian Copula Synthesis (BGCS), a data augmentation method tailored to binary medical data that combines Gaussian copula modeling with a large language model (LLM)-based filtering step to improve the clinical plausibility of generated samples. Second, we evaluate BGCS against established augmentation methods using both synthetic data quality and downstream predictive performance, including binomial proportion tests, dimensionality reduction, and classification experiments across multiple machine learning models. Third, we integrate the best-performing model into an interpretable, decision tree-based clinical decision support system (CDSS) for 90-day early dialysis prediction. To our knowledge, this study is among the first to jointly examine binary-specific data augmentation, synthetic data quality, and interpretable CDSS development for early dialysis prediction in CKD.

% ==========================================
% MATERIALS AND METHODS
% ==========================================
\section*{Materials and methods}

\subsection*{Study design and data source}

This study utilized an EHR dataset obtained from TriNetX, a global federated health research network that provides access to de-identified clinical data from participating healthcare institutions. The study focused on patients diagnosed with CKD in West Virginia, USA. A single-center, retrospective observational design was adopted in accordance with Strengthening the Reporting of Observational Studies in Epidemiology (STROBE) guidelines~\cite{cuschieri2019strobe}.

The study population comprised 90,602 patients whose medical records spanned February 2008 through June 2022. Patients aged 18 years and older diagnosed with CKD using ICD-9-CM code 585 and ICD-10-CM code N18 were included in the study. Dialysis patients were identified using the ICD-10-CM diagnosis code Z99.2 together with dialysis-related CPT (\textit{Current Procedural Terminology}) procedure codes 90935, 90937, 90945, and 90947.

To maintain the analytical focus on the progression toward dialysis initiation, patients who received dialysis before their CKD diagnosis were excluded. Only procedures and medications administered after CKD diagnosis and before the first scheduled dialysis were included. The final dataset comprised 459,868 records from 15,169 unique CKD patients. The detailed data preparation workflow is provided in S1 Appendix.

For each of the independent runs, patients were split into training and testing sets at the patient level. All records from a given patient were assigned exclusively to either the training or testing set to prevent information leakage. Synthetic samples were generated only from the minority-class samples in the training set. The held-out test set was never used during augmentation, LLM filtering, model training, or feature-importance estimation.

\subsection*{Ethics statement}

This study used retrospective, de-identified electronic health record data obtained through the TriNetX platform. Because the dataset was de-identified in accordance with the Health Insurance Portability and Accountability Act (HIPAA), the study qualified for exemption from institutional review board review and informed consent requirements. Accordingly, the study was exempted by the West Virginia University Human Research Review Committee and Institutional Review Board (IRB \#2212689753).

\subsection*{Data characteristics and class imbalance}

In the final analytical dataset (S1 Appendix), approximately 5\% of records belonged to the dialysis class (class 1), indicating marked class imbalance. To support early prediction, data from the 90 days immediately preceding dialysis initiation were excluded so that the model would rely on earlier clinical indicators rather than events occurring close to initiation.

\begin{figure}[!htb]
    \centering
    \includegraphics[scale=0.35]{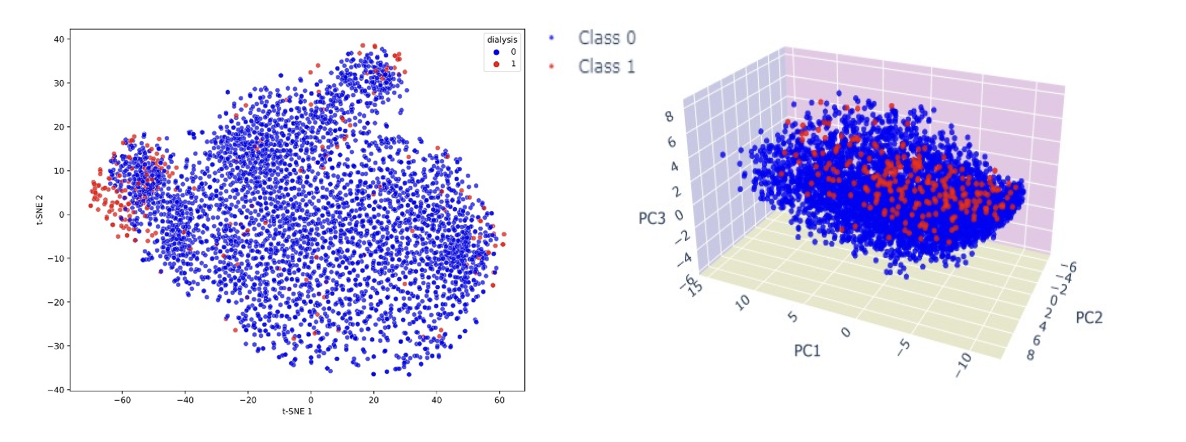}
    \caption{{\bf Visualization of class imbalance in the CKD dataset.}
    Dimensionality reduction using t-SNE (left) and PCA (right) reveals the significant imbalance between dialysis (class 1, approximately 5\%) and non-dialysis (class 0) cases.}
    \label{fig:fig1}
\end{figure}

Fig~\ref{fig:fig1} illustrates the class distribution using t-distributed stochastic neighbor embedding (t-SNE) and principal component analysis (PCA).

\subsection*{Binary Gaussian Copula Synthesis (BGCS)}

Existing augmentation methods such as SMOTE and CTGAN are not specifically designed for binary feature spaces. Given the prevalence of binary data in clinical applications, where diagnoses, procedures, and medications are coded as present or absent, we developed BGCS, a purpose-built augmentation framework for binary medical data.

\subsubsection*{Sampling from the Gaussian copula model.}

BGCS generates synthetic binary samples using a latent Gaussian copula model. 
Let $\mathbf{X} \in \{0,1\}^{N \times p}$ denote the observed binary data 
matrix, where $N$ is the number of records and $p$ is the number of variables. 
BGCS first estimates a $p \times p$ latent correlation matrix $\Sigma$, with 
unit diagonal entries $\Sigma_{jj}=1$, to model dependence among the variables 
within a Gaussian copula framework. In implementation, $\Sigma$ is estimated 
using a regularized latent-correlation estimator for binary variables and 
projected to the nearest positive-definite correlation matrix when necessary. The associated Gaussian copula is defined as

\begin{equation}
C(\mathbf{u}) = \Phi_p\!\left(\Phi^{-1}(u_1), \Phi^{-1}(u_2), \ldots, \Phi^{-1}(u_p); \Sigma\right),
\end{equation}

\noindent where $\Phi_p$ is the cumulative distribution function (CDF) of the 
$p$-variate standard normal distribution with zero mean, unit variances, and 
correlation matrix $\Sigma$; $\Phi^{-1}$ is the quantile function of the 
univariate standard normal distribution; and $\mathbf{u} = (u_1,\ldots,u_p)$ 
with $u_j \in (0,1)$ for all $j$.

To generate one synthetic sample, BGCS draws a latent Gaussian vector

\begin{equation}
\mathbf{Z}^{(t)} = \left(Z_1^{(t)}, Z_2^{(t)}, \ldots, Z_p^{(t)}\right) \sim \mathcal{N}(\mathbf{0}, \Sigma),
\end{equation}

\noindent where $t$ indexes candidate synthetic samples generated sequentially 
from the Gaussian copula. This candidate-generation step is repeated until $m$ samples are retained after LLM-based filtering.

\subsubsection*{Transformation to uniform marginals.}

Each latent Gaussian component is transformed using the standard normal CDF:

\begin{equation}
U_j^{(t)} = \Phi\!\left(Z_j^{(t)}\right), \qquad j = 1,\ldots,p.
\end{equation}

\noindent By construction, each $U_j^{(t)}$ is marginally distributed as Uniform$(0,1)$, while the joint dependence among the components is governed by the Gaussian copula parameterized by $\Sigma$.

\subsubsection*{Binary conversion.}

For each variable $j$, let $\hat{\pi}_j$ denote the empirical marginal probability of observing value 1 in the original data:

\begin{equation}
\hat{\pi}_j = \frac{1}{N}\sum_{r=1}^{N} X_{rj}.
\end{equation}

\noindent Let $\hat{\boldsymbol{\pi}} = (\hat{\pi}_1, \ldots, \hat{\pi}_p)^\top$ 
denote the corresponding vector of empirical marginal probabilities. The synthetic binary sample $\mathbf{S}^{(t)} = (S_1^{(t)},\ldots,S_p^{(t)})$ is then obtained by thresholding the uniform variables:

\begin{equation}
S_j^{(t)} =
\begin{cases}
1 & \text{if } U_j^{(t)} < \hat{\pi}_j,\\
0 & \text{otherwise.}
\end{cases}
\end{equation}

Repeated application of this candidate-generation step produces synthetic binary vectors, which are subsequently screened by the LLM-based filtering stage until the final retained dataset satisfies
\begin{equation}
\mathbf{S} \in \{0,1\}^{m \times p}.
\end{equation}

Equivalently, letting $n_{\mathrm{cand}}$ denote the total number of 
candidates generated before filtering is complete, the copula-based candidate generation step can be written as
\begin{equation}
\tilde{\mathbf{S}} = \mathbb{I}\{\Phi(\mathbf{Z}) < \hat{\boldsymbol{\pi}}\},
\end{equation}
\noindent where $\mathbf{Z} \in \mathbb{R}^{n_{\mathrm{cand}} \times p}$ 
contains $n_{\mathrm{cand}}$ latent Gaussian candidate samples, $\Phi(\mathbf{Z})$ 
is applied elementwise, $\hat{\boldsymbol{\pi}}^\top \in \mathbb{R}^{1\times p}$ is broadcast across the rows of $\Phi(\mathbf{Z})$, and $\mathbb{I}\{\cdot\}$ denotes the elementwise indicator function that equals 1 when its argument is true and 0 otherwise. The matrix $\tilde{\mathbf{S}}$ collects all copula-generated candidates prior to filtering; the final retained dataset $\mathbf{S}$ is obtained by applying the LLM-based filtering stage to the rows of $\tilde{\mathbf{S}}$ until $m$ samples are accepted.

Prior to LLM-based filtering, the marginal probability of each synthetic 
variable matches the corresponding empirical marginal probability by 
construction: 
\begin{equation}
P\!\left(S_j^{(t)} = 1\right) = P\!\left(U_j^{(t)} < \hat{\pi}_j\right) 
= \hat{\pi}_j.
\end{equation}
Because the LLM-based filtering stage preferentially retains samples 
resembling observed minority-class profiles rather than sampling uniformly 
from all candidates, exact marginal preservation is not guaranteed in the 
final retained dataset $\mathbf{S}$. 

Dependence among the synthetic variables is induced through the latent 
Gaussian correlation matrix $\Sigma$. After thresholding, the resulting 
binary samples are intended to approximate the dependence structure of the 
original data, although exact preservation of observed binary pairwise 
dependence is not guaranteed in general. This arises because the nonlinear 
thresholding operation alters the effective correlation in the binary domain 
relative to the latent Gaussian domain. 

\subsubsection*{LLM-based filtering of synthetic samples.}

Although the Gaussian copula model captures dependence through a latent correlation structure, it may still generate binary feature combinations that are statistically plausible under the fitted model but atypical from a clinical perspective. This limitation is especially relevant in high-dimensional EHR data, where diagnosis, procedure, and medication codes may follow patterns involving multiple co-occurring variables that are not fully captured by pairwise dependence alone. Consequently, some copula-generated records may contain uncommon code combinations or may fail to reflect patterns typically observed among real dialysis cases.

To mitigate this issue, BGCS includes a post-generation filtering stage based on a fine-tuned GPT-2 model. The purpose of this stage is not to generate additional samples, but to score the plausibility of copula-generated minority-class candidates based on patterns learned from serialized real dialysis-case records. In this way, the LLM serves as a learned screening model that preferentially retains synthetic samples that more closely resemble observed minority-class profiles.

For each synthetic binary sample $\mathbf{S}^{(t)} = (S_1^{(t)}, S_2^{(t)}, 
\ldots, S_p^{(t)}) \in \{0,1\}^p$, we construct a serialized text 
representation denoted by $T(\mathbf{S}^{(t)})$. This representation maps 
each binary feature to its corresponding feature name in a structured 
sequence, and the sequence is labeled as a minority-class (dialysis) record 
by appending \texttt{Label: 1}. In schematic form, the serialized input is written as

\begin{equation}
T(\mathbf{S}^{(t)}) = \texttt{feature\_1: } S_1^{(t)}, \texttt{ feature\_2: } S_2^{(t)}, \ldots, \texttt{ feature\_p: } S_p^{(t)}, \texttt{ Label: 1.}
\end{equation}

The fine-tuned GPT-2 model is a binary classifier obtained by appending a 
linear classification head to the final hidden state of GPT-2, fine-tuned on 
serialized real dialysis records (Label: 1) and non-dialysis records 
(Label: 0). The serialized sequence $T(\mathbf{S}^{(t)})$ is passed to this model, which produces a scalar logit score $\ell_t \in \mathbb{R}$ from the classification head.  This value is transformed into a plausibility score using the sigmoid function:

\begin{equation}
P_t = \sigma(\ell_t) = \frac{1}{1 + e^{-\ell_t}}.
\end{equation}

Here, larger values of $P_t$ indicate that the synthetic sample is more 
consistent with the minority-class patterns learned from the observed dialysis 
records. Synthetic samples with $P_t > \tau$ are retained, whereas those with 
lower scores are discarded, where $\tau = 0.5$ is the plausibility threshold. The retained samples, therefore, form a filtered synthetic set intended to be more aligned with the empirical structure of the minority class.

This two-stage BGCS procedure combines copula-based generation with LLM-based plausibility screening. The first stage matches empirical marginal probabilities by construction and induces dependence through the latent Gaussian copula model, while the second stage filters generated records using patterns learned from real dialysis cases. Because the copula-based stage is constructed from empirical marginals and a latent correlation structure, this additional filtering step is intended to improve the realism of the synthetic minority-class samples by reducing artifacts that may arise in high-dimensional binary feature spaces. The complete BGCS procedure is summarized in Algorithm~\ref{alg:bgcs}.

\begin{algorithm}[t]
\caption{Binary Gaussian Copula Synthesis (BGCS)}
\label{alg:bgcs}
\scriptsize
\begin{algorithmic}[1]
\Statex \textbf{Inputs:} Minority-class data matrix $\mathbf{X} \in \{0,1\}^{N \times p}$; desired number of retained synthetic samples $m$; fine-tuned GPT-2 classifier $M$; plausibility threshold $\tau = 0.5$
\Statex \textbf{Precompute:} $\hat{\pi}_j = \frac{1}{N}\sum_{r=1}^{N} X_{rj}$ for $j = 1,\ldots,p$
\Statex \hspace{\algorithmicindent} Estimate latent correlation matrix $\Sigma$ from $\mathbf{X}$ using a regularized latent-correlation estimator
\Statex \hspace{\algorithmicindent} Project $\Sigma$ to the nearest positive-definite correlation matrix if necessary
\Statex
\State Initialize empty retained synthetic dataset $\mathcal{S} \gets \emptyset$
\State Initialize retained sample count $n_{\mathrm{acc}} \gets 0$
\State Initialize candidate index $t \gets 1$
\Statex
\While{$n_{\mathrm{acc}} < m$}
    \Statex \textit{\quad // Stage 1: Copula-based generation}
    \State Sample $\mathbf{Z}^{(t)} = (Z_1^{(t)},\ldots,Z_p^{(t)}) \sim \mathcal{N}(\mathbf{0}, \Sigma)$
    \State Compute $U_j^{(t)} = \Phi(Z_j^{(t)})$ for $j = 1,\ldots,p$
    \State Generate candidate sample $S_j^{(t)} = \mathbb{I}(U_j^{(t)} < \hat{\pi}_j)$ for $j = 1,\ldots,p$
    \State Form candidate vector $\mathbf{S}^{(t)} = (S_1^{(t)},\ldots,S_p^{(t)})$
    \Statex \textit{\quad // Stage 2: LLM-based plausibility screening}
    \State Serialize $\mathbf{S}^{(t)}$ as $T(\mathbf{S}^{(t)})$
    \State Pass $T(\mathbf{S}^{(t)})$ to $M$; obtain logit $\ell_t$ from classification head
    \State Compute plausibility score $P_t = \sigma(\ell_t) = \frac{1}{1 + e^{-\ell_t}}$
    \If{$P_t > \tau$}
        \State Append $\mathbf{S}^{(t)}$ to $\mathcal{S}$
        \State $n_{\mathrm{acc}} \gets n_{\mathrm{acc}} + 1$
    \EndIf
    \State $t \gets t + 1$
\EndWhile
\Statex
\State \Return retained synthetic data matrix $\mathbf{S} \in \{0,1\}^{m \times p}$ formed by stacking the samples in $\mathcal{S}$ as rows
\end{algorithmic}
\end{algorithm}

\subsection*{Comparison methods}

BGCS was benchmarked against three established augmentation methods: SMOTE~\cite{chawla2002smote, ghosh2021enhanced}, Conditional Tabular GAN (CTGAN)~\cite{fang2022dp}, and standard Gaussian Copula~\cite{meyer2021copula, lin2021multivariate}. Detailed descriptions are provided in S2 Appendix.

\subsection*{Machine learning classifiers and evaluation}

Four classifiers were used for downstream evaluation: Decision Tree (DT), Random Forest (RF), Logistic Regression (LR), and Support Vector Machine (SVM). All evaluations were conducted over 25 independent runs to ensure robustness.

The evaluation proceeded in three stages: (1) distributional fidelity assessment at the univariate level via cumulative distribution plots and at the multivariate level via binomial proportion tests across all features at six significance levels; (2) structural integrity analysis using PCA and t-SNE to determine whether synthetic data preserved the global and local patterns of the original data; and (3) downstream classification performance, with minority-class recall serving as the primary metric given the clinical imperative to identify patients who will require dialysis.

\subsection*{Performance assessment}

Confusion matrices are fundamental tools in machine learning and statistics, and are primarily used to evaluate the performance of classification models, particularly binary classification models. Typically, the matrix tabulates the number of samples for each combination of the reference class and the predicted class, providing a comprehensive summary of model predictions compared with the true outcomes. In binary classification, it is common to represent the confusion matrix as a positive--negative matrix, where the class of interest is designated as the positive class and the background class is designated as the negative class, as shown in Table~\ref{tab:confusion_matrix}.

A binary classification problem is characterized by four possible outcomes: true positives (TPs), true negatives (TNs), false positives (FPs), and false negatives (FNs). True positives indicate correctly classified positive samples, while true negatives indicate correctly classified negative samples. False positives occur when negative samples are incorrectly classified as positive, whereas false negatives occur when positive samples are incorrectly classified as negative~\cite{powers2011evaluation}.

\begin{table}[!htb]
    \centering
    \caption{{\bf Confusion matrix for binary classification.}
    TP = true positive, TN = true negative, FP = false positive, and FN = false negative.}
    \label{tab:confusion_matrix}
    \begin{tabular}{llcc}
        \toprule
        & & \multicolumn{2}{c}{\textbf{Reference Data}} \\
        \cmidrule(lr){3-4}
        & & \textbf{Positive} & \textbf{Negative} \\
        \midrule
        \multirow{2}{*}{\textbf{Classification Result}}
        & \textbf{Positive} & TP & FP \\
        & \textbf{Negative} & FN & TN \\
        \bottomrule
    \end{tabular}
\end{table}

Several performance metrics can be calculated from this cross-tabulation. Precision, shown in Equation~\eqref{eq:precision}, indicates the proportion of correctly classified positive samples among all samples predicted as positive. Recall, shown in Equation~\eqref{eq:recall}, measures the proportion of positive samples correctly classified among all positive samples in the reference data. Overall accuracy, shown in Equation~\eqref{eq:accuracy}, represents the proportion of correctly classified samples, including both positive and negative classes, among all samples~\cite{powers2011evaluation}.

\begin{equation}
\text{Precision} = \frac{TP}{TP + FP}
\label{eq:precision}
\end{equation}

\begin{equation}
\text{Recall} = \frac{TP}{TP + FN}
\label{eq:recall}
\end{equation}

\begin{equation}
\text{Overall Accuracy} = \frac{TP + TN}{TP + TN + FP + FN}
\label{eq:accuracy}
\end{equation}

Considering the aim of this study, which is the early prediction of dialysis, recall is considered the primary performance metric.
\subsection*{CDSS development}

The best-performing augmentation method was selected for integration into a CDSS. A Decision Tree classifier was chosen as the foundation due to its inherent interpretability, which supports clinician trust and provides transparent, auditable predictions. Feature importance was quantified using a decisiveness metric (S3 Appendix) to identify the clinical variables most influential in dialysis risk stratification.

% ==========================================
% RESULTS
% ==========================================
\section*{Results}

\subsection*{Augmentation fidelity}

\subsubsection*{Univariate feature analysis.}

To assess whether the synthetic datasets preserve the marginal behavior of individual variables, we first conducted a univariate feature analysis. Because the study includes a large number of binary clinical variables, only one representative feature group is presented in the main manuscript for conciseness, while additional feature sets and extended comparisons are reported in the S3 Appendix. Representative variables were selected from groups of related clinical codes based on frequency of occurrence, clinical relevance, and data completeness. These codes were derived from standardized ICD-based diagnosis records and mapped to interpretable clinical conditions during preprocessing. Feature set 1 includes A41.9 (sepsis), D72.829 (elevated white blood cell count), E11.65 (type 2 diabetes mellitus with hyperglycemia), and G89.29 (chronic pain).

\begin{figure}[!htb]
    \centering

    \includegraphics[width=\linewidth]{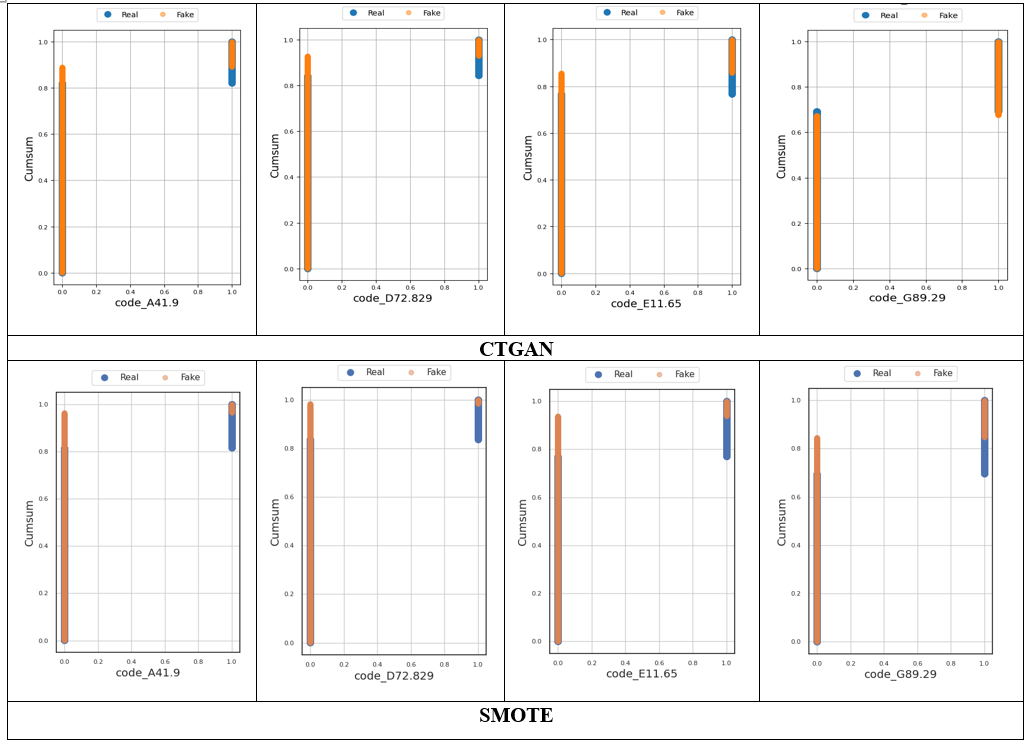}

    \vspace{0.1em}

    \includegraphics[width=\linewidth]{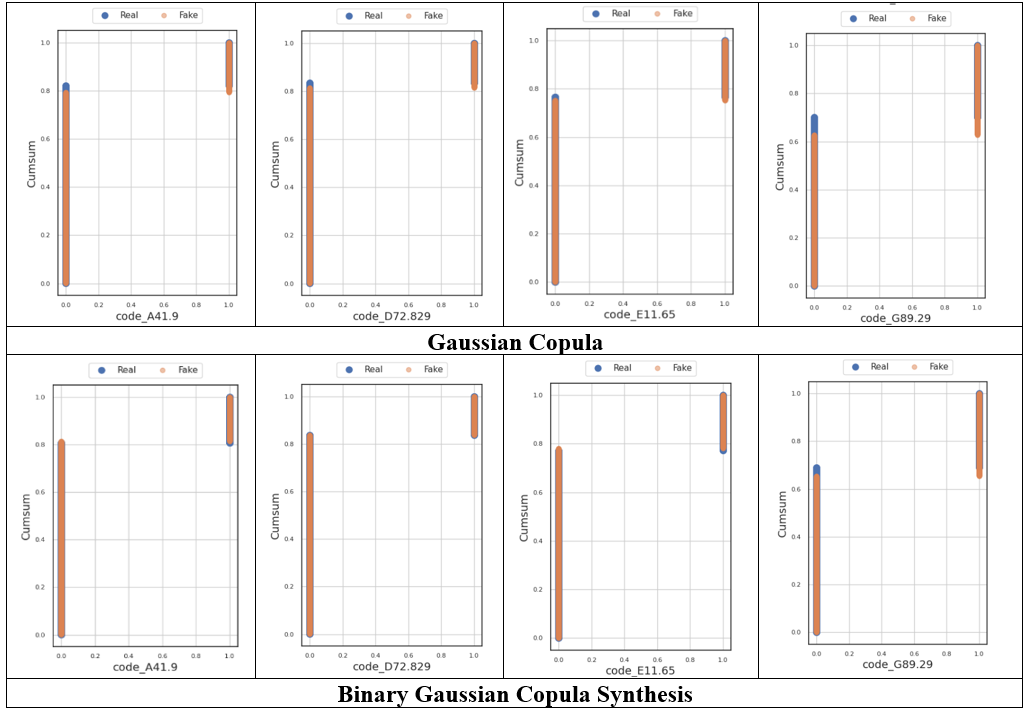}

    \caption{{\bf Cumulative distribution comparison between real and synthetic data for representative clinical features (feature set 1).}
    BGCS and Gaussian Copula most closely reproduce the real data distribution, with BGCS showing marginally higher fidelity.}
    \label{fig:fig2}
\end{figure}

Fig.~\ref{fig:fig2} presents cumulative sum plots comparing the real data with synthetic data generated by each augmentation method. Greater overlap between the curves indicates better preservation of the empirical marginal distribution. The Gaussian Copula and BGCS methods most closely matched the real data across the selected features. Among them, BGCS demonstrated marginally stronger fidelity in reproducing the full cumulative distribution.

% \begin{table}[!htb]
% \centering
% \footnotesize
% \setlength{\tabcolsep}{6pt}
% \renewcommand{\arraystretch}{1.15}
% \caption{{\bf Binomial proportion test results across all features.}
% BGCS achieves the highest mean $p$-value and narrowest Z-statistic range, indicating the closest distributional match to real data.}
% \label{tab:tab2}
% \begin{tabular}{lcccc}
% \toprule
%  & \textbf{SMOTE} & \textbf{CTGAN} & \textbf{Gaussian Copula} & \textbf{BGCS} \\
% \midrule
% \textbf{Z-Statistic Range} 
% & (2.40, 13.47) 
% & ($-$9.21, 6.80) 
% & ($-$2.88, 0.98) 
% & ($-$1.40, 1.31) \\
% \textbf{Z-Statistic Mean} 
% & 6.29 & $-$0.99 & $-$0.39 & $-$0.26 \\
% \textbf{Z-Statistic Std} 
% & 1.82 & 4.30 & 0.71 & \textbf{0.46} \\
% \textbf{P-Value Range} 
% & (2.33E$-$41, 1.62E$-$02) 
% & (3.33E$-$20, 9.96E$-$01) 
% & (3.85E$-$03, 0.99) 
% & \textbf{(0.15, 0.99)} \\
% \textbf{P-value Mean} 
% & 2.23E$-$04 & 8.01E$-$02 & 0.57 & \textbf{0.68} \\
% \textbf{P-value Std} 
% & 1.59E$-$03 & 1.88E$-$01 & 0.26 & 0.19 \\
% \textbf{Dissimilar instances} 
% & 953 & 773 & 97 & \textbf{37} \\
% \textbf{Similar instances} 
% & 7 & 187 & 863 & \textbf{923} \\
% \bottomrule
% \end{tabular}
% \end{table}

\subsubsection*{Statistical analysis.}

A binomial proportion test was applied to compare the marginal proportions of real and synthetic data across the binary clinical features. To keep the main text concise, Table~\ref{tab:tab2} reports aggregate results across all evaluated feature comparisons, while detailed feature-level test results and threshold-specific analyses are provided in the S3 Appendix.

\begin{table}[!htb]
\centering
\footnotesize
\caption{\textbf{Aggregate binomial proportion test results across feature comparisons.}
BGCS achieves the highest mean $p$-value, the narrowest Z-statistic range, and the fewest dissimilar comparisons, suggesting the closest marginal distributional match to the real data.}
\label{tab:tab2}
\resizebox{\linewidth}{!}{%
\begin{tabular}{lcccc}
\toprule
 & \textbf{SMOTE} & \textbf{CTGAN} & \textbf{Gaussian Copula} & \textbf{BGCS} \\
\midrule
\textbf{Z-statistic range} 
& (2.40, 13.47) 
& ($-$9.21, 6.80) 
& ($-$2.88, 0.98) 
& \textbf{($-$1.40, 1.31)} \\
\textbf{Z-statistic mean} 
& 6.29 & $-$0.99 & $-$0.39 & \textbf{$-$0.26} \\
\textbf{Z-statistic SD} 
& 1.82 & 4.30 & 0.71 & \textbf{0.46} \\
\textbf{$p$-value range} 
& (2.33E$-$41, 1.62E$-$02) 
& (3.33E$-$20, 9.96E$-$01) 
& (3.85E$-$03, 0.99) 
& \textbf{(0.15, 0.99)} \\
\textbf{$p$-value mean} 
& 2.23E$-$04 & 8.01E$-$02 & 0.57 & \textbf{0.68} \\
\textbf{$p$-value SD} 
& 1.59E$-$03 & 1.88E$-$01 & 0.26 & 0.19 \\
\textbf{Dissimilar comparisons} 
& 953 & 773 & 97 & \textbf{37} \\
\textbf{Similar comparisons} 
& 7 & 187 & 863 & \textbf{923} \\
\bottomrule
\end{tabular}%
}
\end{table}

SMOTE exhibited a broad Z-statistic range (2.40 to 13.47), reflecting substantial deviation from the real data proportions. CTGAN also showed high variability across features. In contrast, BGCS achieved the narrowest Z-statistic range ($-$1.40 to 1.31), the lowest Z-statistic standard deviation (0.46), and the highest mean $p$-value (0.68), suggesting the closest marginal distributional alignment with the real data. Across the evaluated feature-threshold comparisons, BGCS produced 923 statistically similar comparisons, compared with 863 for Gaussian Copula, 187 for CTGAN, and 7 for SMOTE.

\subsubsection*{Full feature space analysis.}

While univariate analyses assess whether individual feature distributions are preserved, they do not capture whether dependencies among variables and the overall geometry of the dataset remain intact after augmentation. We therefore evaluated the synthetic datasets in the full feature space using dimensionality reduction methods. 

\begin{figure}[!htb]
    \centering
    \includegraphics[width=\linewidth]{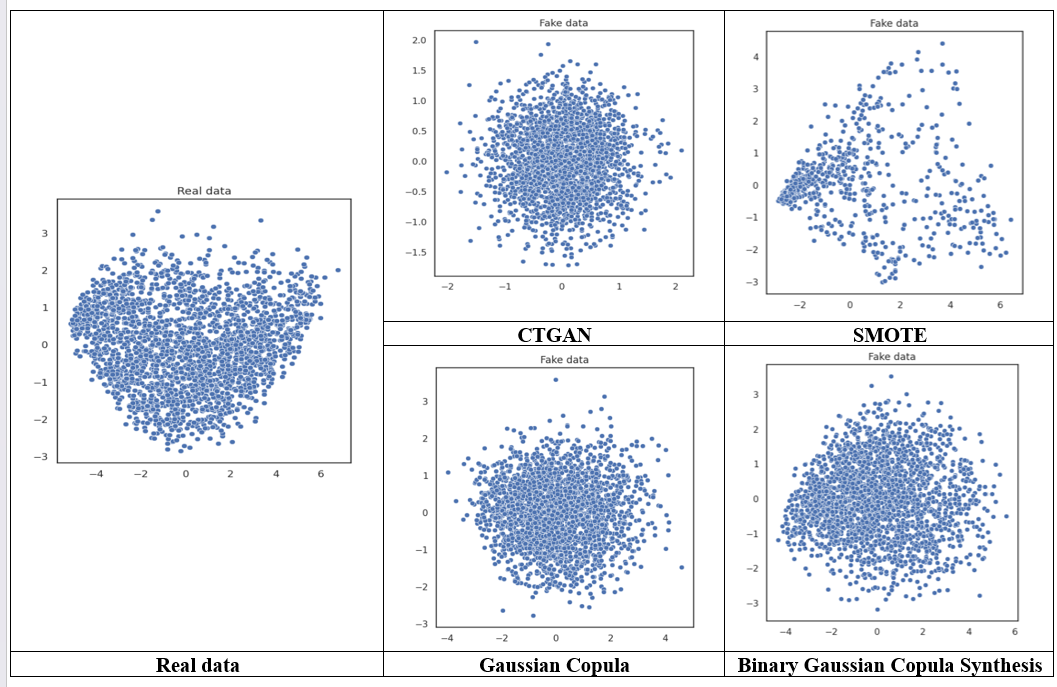}
    \caption{{\bf PCA comparison of real and synthetic data.}
    The first two principal components indicate that BGCS most closely preserves the global structure of the original dataset.}
    \label{fig:fig3}
\end{figure}

Fig.~\ref{fig:fig3} presents principal component analysis (PCA) comparisons between the real and synthetic datasets. PCA summarizes the dominant sources of variation in the high-dimensional feature space into lower-dimensional components, allowing assessment of global structural similarity. Among the evaluated methods, BGCS most closely preserved the overall shape, orientation, and dispersion of the original data cloud, indicating strong retention of large-scale covariance structure.

\begin{figure}[!htb]
    \centering
    \includegraphics[width=\linewidth]{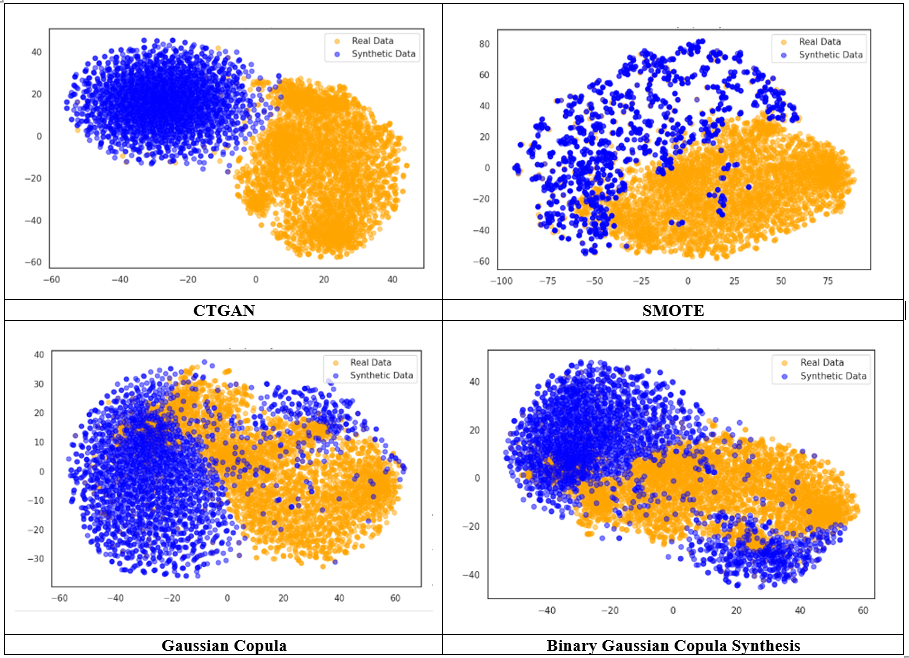}
    \caption{{\bf t-SNE comparison of real and synthetic data (perplexity = 30).}
    BGCS maintains the strongest structural integration with the real data across both local and global scales.}
    \label{fig:fig4}
\end{figure}

To further assess local neighborhood structure and nonlinear relationships, we applied t-SNE with perplexity = 30. As shown in Fig.~\ref{fig:fig4}, BGCS-generated samples exhibited the strongest integration with the real data, with substantial overlap in cluster regions and preservation of both local grouping behavior and broader manifold structure. In contrast, the competing methods showed greater separation or distortion relative to the original data.

Additional visualizations under alternative settings are provided in the S3 Appendix.

\subsection*{Classification performance}

Beyond distributional similarity, the practical value of data augmentation lies in whether it improves downstream predictive performance on the minority class. Because the objective of this study is early identification of patients likely to require dialysis, minority-class recall was treated as the primary evaluation metric. Higher recall indicates that more true dialysis cases are correctly identified. All models were evaluated over 25 independent runs to account for variability arising from data partitioning and model training.

Table~\ref{tab:minority_recall} presents minority-class recall scores across four classifiers for each augmentation method. BGCS consistently achieved the strongest performance, yielding median recall values ranging from 0.78 to 0.87 across models. These results exceeded those obtained from SMOTE, CTGAN, Gaussian Copula, and the unaugmented baseline, indicating that BGCS-generated samples provide more informative decision boundaries for detecting the minority dialysis class.

% Required packages:
% \usepackage{booktabs}
% \usepackage{multirow}
% \usepackage{graphicx}

\begin{table}[!htb]
\centering
\caption{\textbf{Minority-class recall across classifiers and augmentation methods over 25 runs.}
BGCS consistently achieved the highest recall for the dialysis class, with median values ranging from 0.775 to 0.875.}
\label{tab:minority_recall}
\resizebox{\linewidth}{!}{%
\begin{tabular}{llcc}
\toprule
\textbf{Augmentation Method} & \textbf{Classifier} & \textbf{Range} & \textbf{Median} \\
\midrule
\multirow{4}{*}{CTGAN}
    & Decision Tree       & 0.525--0.675 & 0.575 \\
    & Random Forest       & 0.525--0.650 & 0.600 \\
    & SVM                 & 0.600--0.650 & 0.625 \\
    & Logistic Regression & 0.600--0.750 & 0.650 \\
\midrule
\multirow{4}{*}{SMOTE}
    & Decision Tree       & 0.500--0.700 & 0.575 \\
    & Random Forest       & 0.625--0.700 & 0.675 \\
    & SVM                 & 0.675--0.750 & 0.725 \\
    & Logistic Regression & 0.650--0.700 & 0.675 \\
\midrule
\multirow{4}{*}{Gaussian Copula}
    & Decision Tree       & 0.600--0.800 & 0.725 \\
    & Random Forest       & 0.700--0.775 & 0.750 \\
    & SVM                 & 0.675--0.775 & 0.725 \\
    & Logistic Regression & 0.700--0.825 & 0.775 \\
\midrule
\multirow{4}{*}{BGCS}
    & Decision Tree       & 0.625--0.875 & 0.775 \\
    & Random Forest       & 0.775--0.825 & 0.800 \\
    & SVM                 & 0.800--0.900 & 0.850 \\
    & Logistic Regression & 0.800--0.925 & 0.875 \\
\bottomrule
\end{tabular}%
}
\end{table}

The gains were observed across all four classifiers, suggesting that the advantage of BGCS is not model-specific but instead reflects improved representation of the minority-class feature space. This consistency is particularly important in clinical prediction settings, where robust performance across modeling frameworks strengthens confidence in the augmentation strategy.

\begin{figure}[!htb]

    \includegraphics[width=\linewidth]{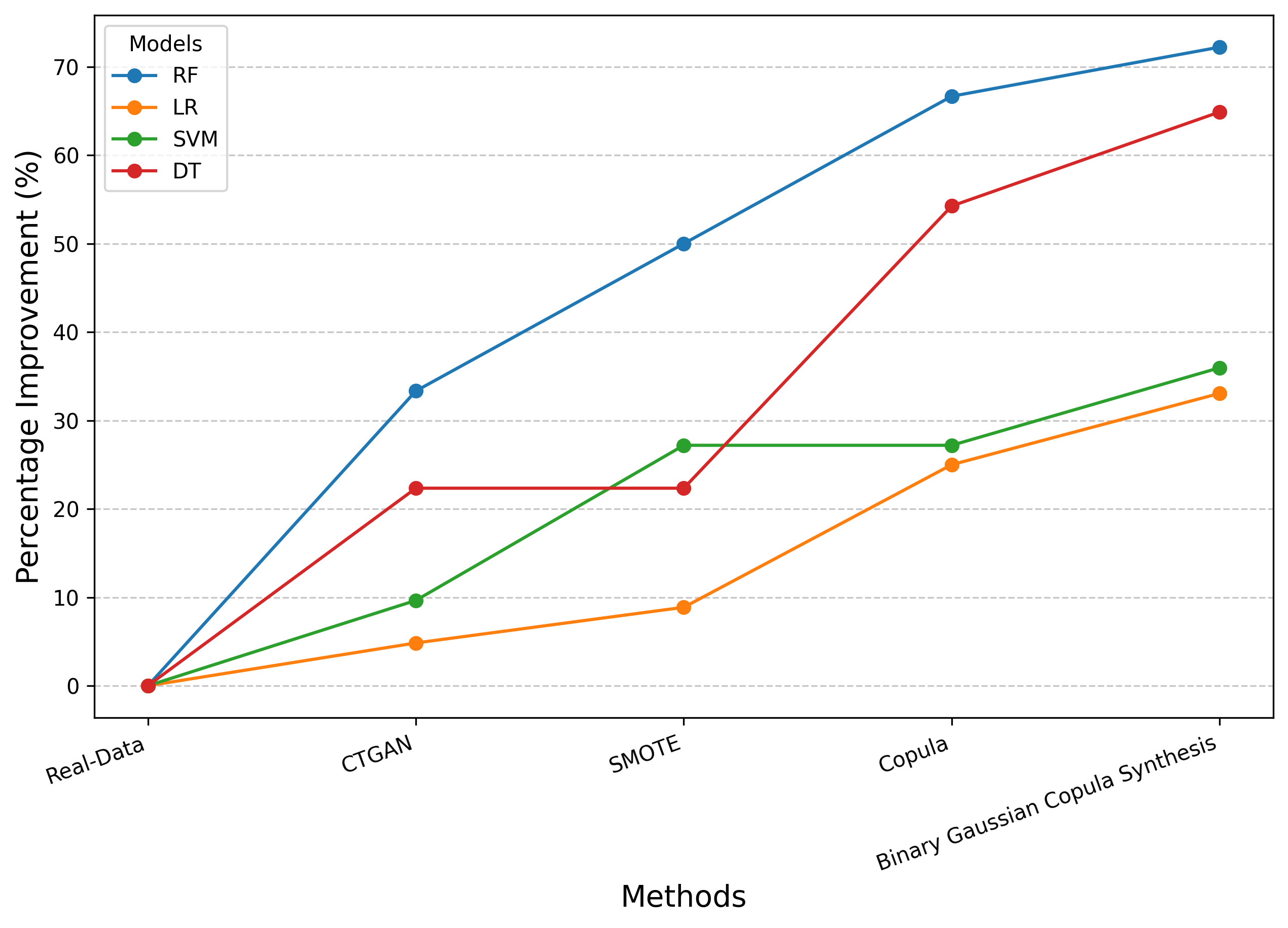}

    \caption{{\bf Recall comparison between models trained on augmented and real data.}
 The recall improvement achieved by each augmentation method relative to real data.}
    \label{fig:fig6}
\end{figure}

For the top-performing models, BGCS improved recall by approximately 72\% for Random Forest and 65\% for Decision Tree relative to training on real data alone. Fig.~\ref{fig:fig6} compares recall for models trained on augmented and unaugmented data, illustrating the magnitude of improvement relative to using the original imbalanced dataset alone. Across classifiers, BGCS produced the largest and most consistent recall gains.

\subsection*{Clinical decision support system}

Given the consistent superiority of BGCS in both distributional fidelity and downstream recall, BGCS-augmented datasets were selected for CDSS development. Because interpretability is essential in clinical settings, a Decision Tree model was used as the foundation of the CDSS. As shown in Table~\ref{tab:minority_recall} and Fig~\ref{fig:fig6}, the BGCS-augmented Decision Tree substantially improved minority-class recall relative to the unaugmented Decision Tree.

\begin{figure}[!htb]
    \centering
    \includegraphics[scale=0.8]{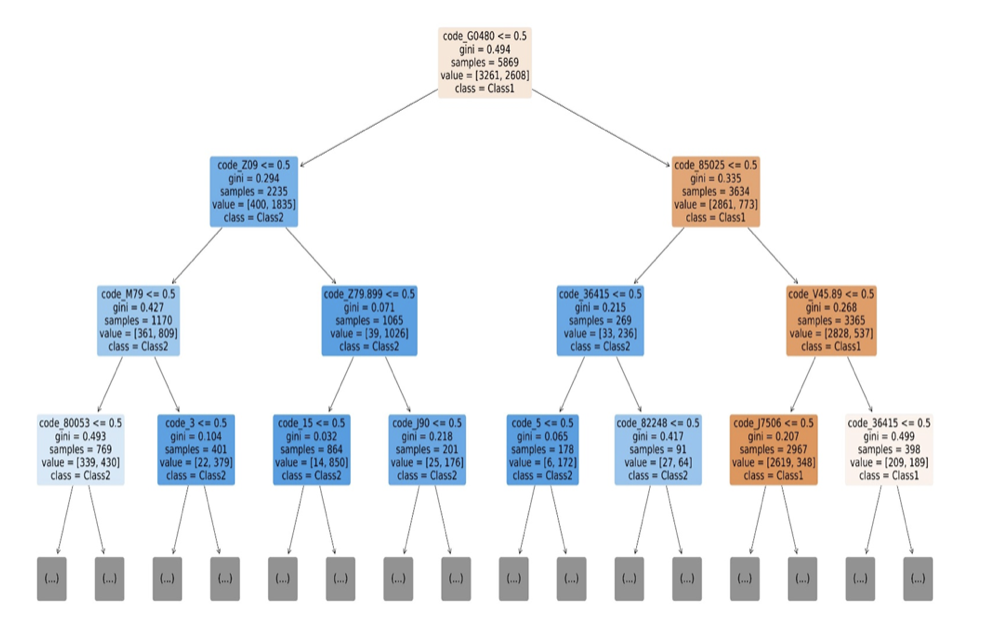}
    \caption{{\bf Truncated decision tree for the BGCS-augmented CDSS.}
    The hierarchical arrangement of clinical features illustrates the branching logic used for dialysis risk stratification.}
    \label{fig:fig7}
\end{figure}

Fig.~\ref{fig:fig7} depicts a truncated decision tree highlighting the hierarchical positioning of features in the selected model. The root node, ``code\_G0480,'' corresponds to drug identification methods for individual drugs and structural isomers. At the second level, ``code\_Z09'' represents follow-up examination for conditions other than malignant neoplasm, and ``code\_85025'' represents complete blood count.

Features near the top of the tree are useful for early data separation, but they are not necessarily the most influential variables overall. Therefore, Table~\ref{tab:top10} lists the top 10 influential features identified from the fitted Decision Tree using impurity-based feature importance, measured by the mean decrease in Gini impurity. This index quantifies the cumulative contribution of each feature to reducing class impurity and separating dialysis from non-dialysis cases across all splits in the tree. These codes correspond to clinical conditions, laboratory tests, and physiological markers associated with kidney function deterioration and dialysis risk.

\begin{table}[!htb]
\centering
\footnotesize
\setlength{\tabcolsep}{6pt}
\renewcommand{\arraystretch}{1.25}
\caption{{\bf Top 10 influential clinical features for dialysis prediction.}
Features identified by the Decision Tree model as important for distinguishing dialysis from non-dialysis outcomes, with corresponding ICD-10-CM or CPT codes and clinical interpretations.}
\label{tab:top10}
\begin{tabular}{cll}
\toprule
\textbf{Rank} & \textbf{Code} & \textbf{Clinical interpretation} \\
\midrule
1  & E87.5  & Hyperkalemia \\
2  & E87.6  & Hypokalemia \\
3  & I25.2  & Myocardial infarction \\
4  & I10    & Hypertension \\
5  & 81001  & Urinalysis by dipstick or tablet reagent \\
6  & 84100  & Phosphate level in patient specimen \\
7  & I11.0  & Hypertensive heart disease with heart failure \\
8  & 84295  & Sodium concentration \\
9  & I51.7  & Cardiomegaly \\
10 & R94.31 & Nonspecific abnormal ECG/EKG \\
\bottomrule
\end{tabular}
\end{table}

\noindent The most influential features reflect electrolyte imbalances, including hyperkalemia, hypokalemia, sodium levels, and phosphate levels; cardiovascular complications, including hypertension, myocardial infarction, hypertensive heart disease, and cardiomegaly; and diagnostic monitoring indicators, including urinalysis and abnormal ECG/EKG findings. These variables are clinically consistent with dialysis risk in CKD, where electrolyte dysregulation and cardiovascular comorbidities are important markers of disease progression.

\begin{figure}[!htb]
    \centering
    \includegraphics{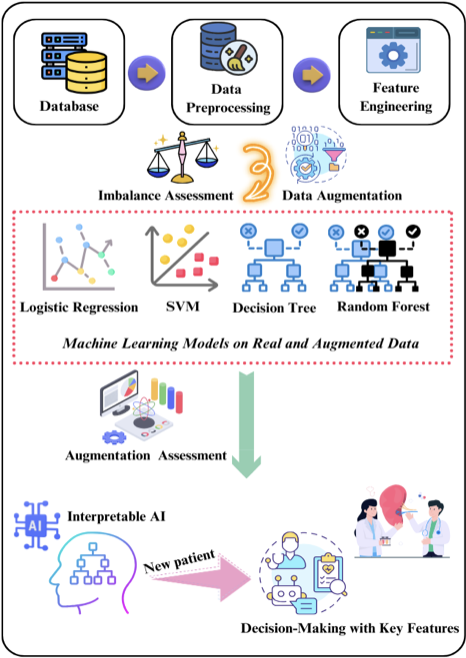}
    \caption{{\bf Overall framework of the proposed CDSS.}
    The system integrates data preprocessing, BGCS augmentation, classifier training, and interpretable prediction to support early dialysis risk stratification.}
    \label{fig:fig8}
\end{figure}

The overall CDSS framework is illustrated in Fig.~\ref{fig:fig8}, showing how data preparation, BGCS augmentation, model training, and interpretable prediction are integrated into a unified pipeline.

The proposed CDSS is designed as an auxiliary tool to complement clinical judgment. Its predictions are intended to be interpreted alongside clinical expertise, providing timely, data-driven insights to support early dialysis risk stratification and proactive care planning.

% ==========================================
% DISCUSSION
% ==========================================
\section*{Discussion}

This study presents BGCS, a data augmentation method designed specifically for binary clinical data, and demonstrates its utility within an interpretable CDSS for early dialysis prediction in CKD patients. Our results show that BGCS consistently outperforms established augmentation methods across distributional fidelity, structural integrity, and downstream classification performance.

The central finding is that augmentation methods tailored to the structure of clinical data can substantially improve predictive performance for rare but critical health outcomes. BGCS achieved the highest distributional similarity to real data (mean $p$-value of 0.68 across all features), while SMOTE and CTGAN showed significantly greater deviations. This advantage likely stems from two design features: the Gaussian copula framework explicitly models pairwise dependencies among binary variables, preserving the correlation structure of the original data; and the LLM-based filtering step removes statistically valid but clinically implausible samples, improving the quality of augmented data.

The downstream impact was substantial. Across all four classifiers, BGCS-augmented datasets yielded the highest minority-class recall (median 0.78--0.87). The consistency of this improvement across diverse classifier architectures---from interpretable models (Decision Tree, Logistic Regression) to more complex ones (Random Forest, SVM)---suggests that the benefit is attributable to the quality of the augmented data rather than to a favorable interaction with a specific model.

These findings extend prior work on class imbalance in healthcare. Several studies have applied SMOTE and related techniques to CKD datasets~\cite{razzaghi2019predictive, shi2022resampling}, but these methods were not designed for binary data and, as our results confirm, produce synthetic samples with substantial distributional deviations. More recent approaches using GANs and VAEs~\cite{nicholas2023generating, weng2024joint} have shown promise for tabular clinical data but remain limited for purely binary EHR datasets. BGCS addresses this gap by combining copula-based statistical modeling with LLM-driven quality filtering.

The integration of BGCS into a decision tree--based CDSS reflects the growing emphasis on interpretability in clinical AI~\cite{du2022explainable, jeong2022pivotal}. The decision tree structure provides transparent, auditable reasoning that clinicians can inspect and verify, addressing a well-documented barrier to adoption of machine learning in clinical settings. The identification of decisive clinical features offers actionable insights for risk stratification and can inform the timing and intensity of monitoring for high-risk patients.

\subsection*{Clinical implications}

Beyond technical performance, the clinical significance of these findings warrants careful consideration. In CKD management, a false negative, namely a patient who will require dialysis but is not identified by the model, may carry serious consequences. Patients who progress to dialysis without adequate preparation are substantially more likely to initiate treatment through emergency hemodialysis using temporary central venous catheters rather than through planned arteriovenous fistula (AVF) creation~\cite{shimizu2020emergent}. Because AVF creation often requires several months for surgical placement and maturation~\cite{lok2020kdoqi}, delayed identification can bypass this critical preparation window. Emergency dialysis starts have been associated with higher mortality, greater infection risk, increased hospitalization, and higher healthcare costs~\cite{shimizu2020emergent,michel2018deleterious}. The recall values of 0.78 to 0.87 achieved by BGCS-augmented models suggest that approximately 8 of every 10 future dialysis patients could be identified in advance under the evaluated setting. Even modest gains in earlier identification may expand opportunities for vascular access planning, patient education, and consideration of pre-emptive transplantation.

From a workflow perspective, the proposed CDSS is intended to function as a passive screening layer within existing EHR infrastructure. Rather than replacing clinical judgment, the system would flag patients whose binary clinical profiles indicate elevated dialysis risk, prompting nephrology review while intervention opportunities remain. The decision tree architecture is well suited to this role because its transparent branching logic allows clinicians to inspect the rationale behind each alert. In practice, the system could support accelerated referrals for vascular access evaluation, dietary counseling for electrolyte management, and intensified cardiovascular monitoring among flagged patients. Workflow-compatible integration of this type has been identified as an important factor in successful adoption of ML tools in healthcare settings~\cite{kawamoto2005improving}.

It is important to acknowledge that retrospective predictive performance does not guarantee clinical utility. Translation from model recall to patient benefit depends on factors not evaluated here, including clinician acceptance, alert fatigue, operating thresholds, specificity, and the availability of resources to act on early predictions. Prospective studies examining clinical decision-making, time to vascular access creation, and rates of emergency versus planned dialysis initiation are needed to determine whether improved predictive performance translates into meaningful patient outcomes~\cite{du2022explainable,jeong2022pivotal}.

Importantly, the recall values reported in this study were obtained without hyperparameter tuning; classifiers were trained using default configurations. This design choice was deliberate, because holding model settings constant across augmentation strategies helps isolate the effect of data quality rather than classifier-specific optimization. From a clinical standpoint, the reported detection rates may therefore represent a conservative estimate. Systematic hyperparameter optimization may yield further improvements and could further reduce missed high-risk patients. This suggests that combining BGCS augmentation with optimized classifiers is a promising direction for future prospective nephrology deployment.

\subsection*{Limitations}

Several limitations should be acknowledged. First, the dataset was drawn from a single geographic region, West Virginia, which may limit generalizability. West Virginia has among the highest CKD burdens in the United States, and its patient population may differ demographically and clinically from populations in other healthcare systems. External validation using more geographically and demographically diverse datasets is therefore needed.

Second, the LLM-based filtering step introduces dependence on the fine-tuned language model. The effectiveness of this step may vary with the training data, clinical domain, and representation of patient records. Future work should evaluate the sensitivity of BGCS to different language models and assess whether the filtering step remains clinically consistent across external datasets.

Third, the proposed CDSS was evaluated retrospectively. Prospective evaluation is needed to assess clinician acceptance, workflow integration, alert burden, and impact on patient outcomes, including earlier vascular access planning and reduced emergency dialysis initiation.

Finally, although the Decision Tree classifier was selected for interpretability, it may sacrifice predictive performance relative to ensemble or deep learning models. Future studies should investigate hybrid approaches that combine the transparency of interpretable models with the predictive strength of more flexible classifiers.

\subsection*{Future directions}

Future work should validate BGCS across larger, multi-center, and demographically diverse EHR datasets to assess robustness and transportability. Additional distributional similarity measures, such as kernel-based distances, maximum mean discrepancy, or Wasserstein distance, could provide more nuanced evaluation of synthetic data quality beyond marginal proportion tests and dimensionality reduction.

Methodologically, hybrid augmentation strategies that combine copula-based dependence modeling with deep generative models may further improve representation of complex clinical patterns. Future studies should also examine the sensitivity of BGCS to different LLM architectures and filtering strategies. From a clinical translation perspective, prospective deployment studies are needed to determine whether the proposed CDSS improves clinical decision-making, vascular access planning, emergency dialysis reduction, and patient outcomes.

\section*{Acknowledgments}

We would like to express our appreciation to the Department of Industrial and Management Systems Engineering (IMSE), West Virginia University, for their support during this research project.

\subsection*{Data availability}

The data underlying this study were obtained from the TriNetX federated health research network under a data use agreement and cannot be made publicly available. Researchers interested in accessing TriNetX data may submit an application through their institution via the TriNetX platform at \url{https://trinetx.com}. All other materials, code, and analytical outputs supporting the findings of this study are available from the corresponding author upon reasonable request.

\bibliography{references}

% ==========================================
% SUPPORTING INFORMATION
% ==========================================
\section*{Supporting information}

\paragraph*{S1 Appendix.}
\label{S1_Appendix}
{\bf Dataset description and preparation.} Data preparation workflow including missing
value handling, feature engineering, and early prediction filtering. A complete list of clinical features used in the study with corresponding ICD-10-CM codes are available in this section, too.

\paragraph*{S2 Appendix.}
\label{S2_Appendix}
{\bf Comparison augmentation methods} Detailed descriptions of CTGAN and SMOTE augmentation methods are included in this section.

\paragraph*{S3 Appendix.}
\label{S3_Appendix}
{\bf Extended results.} Additional univariate feature analysis across feature sets~2 and~3 with probability distribution and violin plots; per-feature binomial proportion test results for each augmentation method; additional full feature space comparisons using t-SNE at low and high perplexity settings and log-scale mean comparisons.

\end{document}